\newcommand{\CLASSINPUTtoptextmargin}{2cm}
\newcommand{\CLASSINPUTbottomtextmargin}{2.6cm}
\documentclass[conference]{IEEEtran}
\IEEEoverridecommandlockouts

\usepackage{cite}
\usepackage{amsmath,amssymb,amsfonts}
\usepackage{graphicx}
\usepackage{textcomp}
\usepackage{xcolor}

\usepackage{algorithmicx}
\usepackage{algorithm}
\usepackage{algpseudocode, graphicx}

\usepackage{xcolor}

\def\BibTeX{{\rm B\kern-.05em{\sc i\kern-.025em b}\kern-.08em
    T\kern-.1667em\lower.7ex\hbox{E}\kern-.125emX}}
\begin{document}

\title{Joint Deployment and Beamforming Design of Aerial STAR-RIS Aided Networks with Reinforcement Learning\\
 \thanks{This work is partly supported by the Guangdong University Young Innovative Talents Program Project under Grant No.2024KQNCX050, National Natural Science Foundation of China under Grant No.72401122, Guangdong Basic and Applied Basic Research Foundation under Grants No.2024A1515012241 \textit{(Corresponding author: Bai Yan.)}}

}





\author{
\IEEEauthorblockN{Zhuoyuan Ma$^{1,2}$, Qi Zhao$^{1}$, Jin Zhang$^{1,2}$, Bai Yan$^{3}$}
\IEEEauthorblockA{$^1$ Department of Computer Science and Engineering, Southern University of Science and Technology, Shenzhen, China}
\IEEEauthorblockA{$^2$ Research Institute of Trustworthy Autonomous Systems, Southern University of Science and Technology, Shenzhen, China}
\IEEEauthorblockA{$^3$ School of Computer Science and Technology, Dongguan University of Technology, Dongguan, China}
\IEEEauthorblockA{mazy2023@mail.sustech.edu.cn, \{zhaoq, zhangj4\}@sustech.edu.cn, yanbai@dgut.edu.cn}
}


\maketitle

\begin{abstract}
Aerial simultaneous transmitting and reflecting reconfigurable intelligent surfaces (STAR-RIS) enables full-space coverage in dynamic wireless networks. However, most existing works assume fixed user grouping, overlooking the fact that STAR-RIS deployment inherently determines whether users are served via transmission or reflection.
To address this, we propose a joint deployment and beamforming framework, where an aerial STAR-RIS dynamically adjusts its location and orientation to adaptively control user grouping and enhance hybrid beamforming. We formulate a Markov decision process (MDP) capturing the coupling among deployment, grouping, and signal design.
To solve the resulting non-convex and time-varying problem, we develop a PPO-based reinforcement learning algorithm that adaptively balances user grouping and beamforming resources through online policy learning. Simulation results show 57.1\% and 285\% sum-rate gains over fixed-deployment and RIS-free baselines, respectively, demonstrating the benefit of user-grouping-aware control in STAR-RIS-aided systems.
\end{abstract}


\section{Introduction}
\IEEEPARstart{R}{econfigurable} intelligent surfaces (RISs) have been recently proposed as a promising technology to reconfigure a propagation environment of wireless communication\cite{2inSTARZhong,3inSTARZhong}. Conventional RISs either transmit or reflect the incident signals, merely achieving half-space coverage. Simultaneous transmitting and reflecting RISs (STAR-RISs) overcome this drawback \cite{12inSTARZhong}, providing a full-space smart radio environment. 

There have been several works investigating STAR-RIS's benefits to a variety of communication systems. Most of them optimize the passive beamforming of STAR-RIS, aiming to enhance communication rate \cite{nonperfectCSI, ZhongRuikang, YAN2022109725}, enlarge coverage range \cite{multihop}, or reduce energy consumption \cite{usermovements, UAVenergy}.
These works employ a fixed deployment scheme for the STAR-RIS, where the location and orientation of the STAR-RIS are unable to change. This apparently restricts the wireless communication system performance. 

To remit the restriction, some efforts have been made on optimizing the location, orientation, or both of them. The deployment location and orientation of STAR-RIS co-determine the \textit{user grouping}. The grouping in turn determines whether a user's signal is served by the transmission or reflection beamforming of STAR-RIS. In work \cite{orientation}, STAR-RIS orientation is  optimized with passive beamforming, but the deployment location is unexplored. Some works \cite{gao2022joint,wang2023average,xiao2024star} focus on finding an optimal location but with an implicit assumption of fixed user grouping. Further works consider either a few user-grouping configurations \cite{zhang2022joint,su2023joint,aung2023aerial} or static scenarios that user locations and channel conditions remain constant over time \cite{orideploy, new_dep, 2025static}. They fail to explore joint location and orientation schemes for realistic mobile communication systems where user locations and channel conditions evolve over time.

The above existing works on dynamic STAR-RIS often overlook or oversimplify its impact on user grouping, limiting the ability to adapt transmission and reflection modes in dynamic environments. To overcome this, we propose an aerial STAR-RIS-aided multiple-input single-output (MISO) system where the STAR-RIS, mounted on a UAV, can adapt its location and orientation in real time. This dynamic capability allows for optimized user grouping and enhanced system performance. Specifically, this work contributes the following:

\begin{itemize}
    \item We introduce the first dynamic deployment scheme enabling simultaneous, real-time optimization of both STAR-RIS location and orientation, specifically addressing realistic mobile communication systems where user locations and channel conditions continuously change. 
    
    \item We formulate a joint optimization problem maximizing sum rate by capturing the interplay between STAR-RIS mobility and hybrid beamforming.
    
    \item We develop a PPO-based reinforcement learning algorithm to solve this problem. Given the problem's non-convexity, dynamic nature, and real-time requirements, which render traditional optimization methods insufficient, our proposed algorithm efficiently manages the joint dynamic deployment and hybrid beamforming.
\end{itemize}
Simulation results show that the proposed scheme consistently outperforms both fixed deployment and STAR-RIS-free scenarios, achieving improvements of 57.1\% and 285\% in the sum rate, respectively. These results clearly demonstrate the superior performance of the proposed scheme in addressing the challenges of realistic dynamic mobile communication systems.

The remainder of this paper is organized as follows: Section II details the system model and problem formulation. Section III presents the PPO-based algorithm. Section IV discusses the simulation results, and Section V concludes the paper.

\section{System Model And Problem Formulation}
\subsection{System Model}
\begin{figure*}[t]
    \centering
        \includegraphics[width=0.8\textwidth]{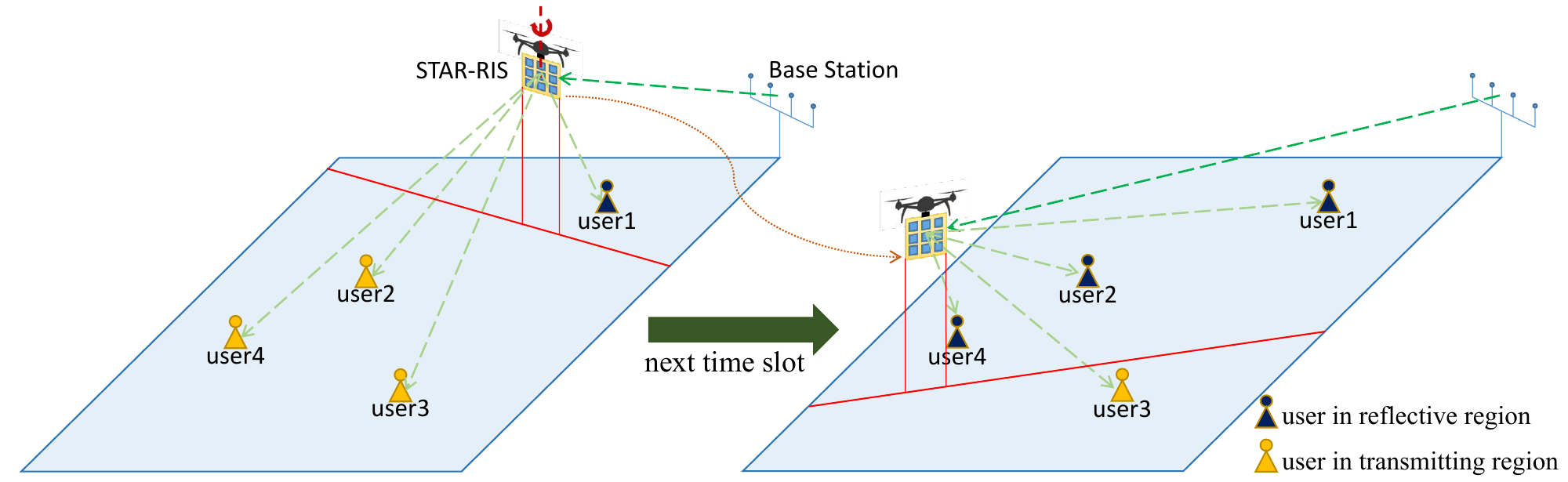}
    \centering
        \caption{Aerial STAR-RIS assisted wireless communication system}
    \label{system model}
\end{figure*}

As shown in Fig. \ref{system model}, we consider an aerial STAR-RIS-aided MISO downlink network, where a BS with linearly placed $M$ antennas serves $K$ single-antenna users, with the transmission and reflection functionality of a STAR-RIS. The STAR-RIS is a uniform planar array with $N$ elements and is mounted on a UAV to dynamically adjust the location and orientation to boost the system performance. 

\begin{figure}[t]
\centering
\includegraphics[width=0.45\textwidth]{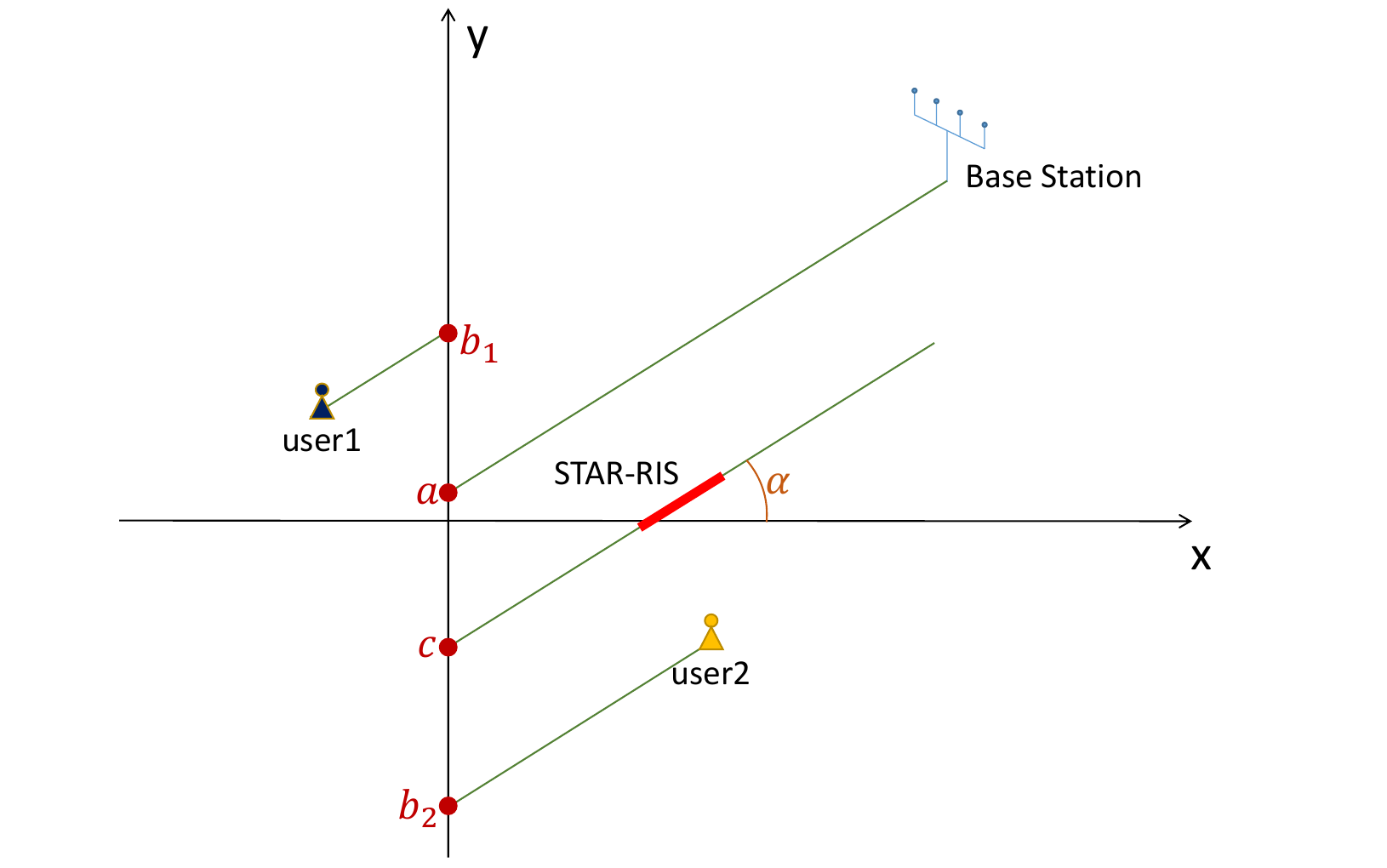}
\caption{User grouping function demonstration}
\label{func}
\end{figure}

\textbf{User Grouping Geometry Model:} The deployment location and orientation of STAR-RIS codetermine the user grouping. 
To elaborate user grouping mathematically, we assume the location of the BS, STAR-RIS, and user $u_k$ as $(x_{B}, y_{B}, z_{B})$, $(x_{R}, y_{R}, z_{R})$, and $(x_k, y_k, z_k)$, respectively. 
The STAR-RIS is perpendicular to the ground and rotates solely along the global z-axis with Euler angle $\alpha$ between its surface and x-axis, yielding rotation matrix $\mathbf{R}_R$ according to \cite{tutorial}. 
Within the two-dimensional x-y plane depicted in Fig. \ref{func}, lines intersecting BS, $u_k$ and STAR-RIS at angle $\alpha$ generate y-axis intercepts $a$, $b_k$, $c$ respectively: 
\begin{equation}
\label{region}
\begin{aligned}
f(u_k) &= (a-c) \times (b_k - c), \\
\text{with} \quad
a &= \sin\alpha \times x_{B} - \cos\alpha \times y_{B}, \\
b_k &= \sin\alpha \times x_{k} - \cos\alpha \times y_{k}, \\
c &= \sin\alpha \times x_{R} - \cos\alpha \times y_{R}. 
\end{aligned}
\end{equation}
We can infer that, if $f(u_k)>0$, the BS and user $u_k$ are on the same sides of the STAR-RIS, i.e., $u_k$ is in reflective region; otherwise, the $u_k$ is in transmitting region. 
We prohibit the occurrence of $f(u_k)=0$ due to its lack of significance, a situation that can be readily mitigated through UAV control.

\subsubsection{\textbf{STAR-RIS Basis}}
We employ the Energy Splitting (ES) protocol  \cite{12inSTARZhong} for the STAR-RIS. The reflected and transmitted signals of the $n$-th STAR-RIS element are denoted as $\beta_{R,n}e^{j\theta_{R,n}}$ and $\beta_{T,n}e^{j\theta_{T,n}}$ respectively, where $n\in\mathcal{N}\triangleq\{1, 2,\ldots, N\}$. Moreover, due to the conservation of energy law\cite{14inSTARZhong}, the reflection phase shift, transmission phase shift, reflection amplitude, and transmission amplitude of the STAR-RIS signals follow the equation:

\begin{equation}
\label{limit}
\begin{aligned}
    &\beta_{R,n}\beta_{T,n}cos(\theta_{R,n}-\theta_{T,n})=0, \\
    &\beta_{T,n}^2+\beta_{R,n}^2=1.
\end{aligned}
\end{equation}

Therefore, to calculate the received signal of user $u_k$ from STAR-RIS, we represent the passive beamforming between STAR-RIS and user $u_k$ as a diagonal matrix:
\begin{equation}
\label{diagmatrix}
\begin{aligned}
\centering
\Theta_{k}\!\triangleq\! &
    \begin{cases}
    \!\Theta_{R}\!=\!diag(\beta_{R,1}e^{j\theta_{R,1}}\!,\!\ldots,\!\beta_{R,N}e^{j\theta_{R,N}}\!), & \text{if } f(u_k) > 0, \\
    \!\Theta_{T}\!=\!diag(\beta_{T,1}e^{j\theta_{T,1}}\!,\!\ldots,\!\beta_{T,N}e^{j\theta_{T,N}}\!), & \text{otherwise}.
    \end{cases}
\end{aligned}
\end{equation}

\subsubsection{\textbf{Channel Model}}
We assume all channels follow quasi-static fading model, and perfect channel state information (CSI) can be obtained by efficient channel estimation technique \cite{ChannelEstimation} under ES protocol. 
In STAR-RIS's local coordinate system (LCS), the angle-of-arrival (AOA) from BS consists of an azimuth angle and an elevation angle, i.e., $[\phi^A_R, \psi^A_R]$:
\begin{equation}
    \phi^A_R = \arctan2([\mathbf{R}_R^T(\mathbf{p}_R-\mathbf{p}_B)]_2, [\mathbf{R}_R^T(\mathbf{p}_R-\mathbf{p}_B)]_1),
\end{equation}
\begin{equation}
    \psi^A_R = \arcsin([\mathbf{R}_R^T(\mathbf{p}_R-\mathbf{p}_B)]_3 / ||\mathbf{p}_R - \mathbf{p}_B||_2),
\end{equation}
and STAR-RIS angle-of-departure (AOD) $[\phi^D_{Rk}, \psi^D_{Rk}]$ towards user $u_k$ is:
\begin{equation}
    \phi^D_{Rk} = \arctan2([\mathbf{R}_R^T(\mathbf{p}_{u_k}-\mathbf{p}_R)]_2, [\mathbf{R}_R^T(\mathbf{p}_{u_k}-\mathbf{p}_R)]_1),
\end{equation}
\begin{equation}
    \psi^D_{Rk} = \arcsin([\mathbf{R}_R^T(\mathbf{p}_{u_k}-\mathbf{p}_R)]_3 / ||\mathbf{p}_{u_k} - \mathbf{p}_R||_2).
\end{equation}

Analogously, in the BS’s LCS, the AOD of BS $[\phi^D_B, \psi^D_B]$ towards STAR-RIS is calculated with a similar method towards $[\phi^D_{Rk}, \psi^D_{Rk}]$.

The element spacing of BS and STAR-RIS are denoted as $d_a$ and $d_e$ respectively, and both are half of signals wavelength. The array response vectors of the $m$-th BS antenna and the $n$-th STAR-RIS element are denoted as $\mathbf{a}_b\in \mathbb{C}^{M \times 1}$ and  $\mathbf{a}_r\in \mathbb{C}^{N \times 1}$:
\begin{equation}
    \mathbf{a}_b[m] = e^{j2\pi(m-1)d_asin(\phi^D_B)/\lambda},
\end{equation}
\begin{equation}
\begin{aligned}
    \mathbf{a}_r[n] &= e^{j2\pi(n-1)d_e(\lfloor n/N_x \rfloor cos(\phi))} \\ 
        &\quad \times e^{(n-\lfloor n/N_x \rfloor N_x) sin(\phi)cos(\psi) / \lambda},
\end{aligned}
\end{equation}
where $N_x$ denotes the number of STAR elements in each row and $\lambda$ is the wavelength of signals.\par

Both BS and the STAR-RIS have a Line-of-Sight (LoS) component due to their respective positions, and STAR-RIS creates virtual LoS between BS and each user. Therefore, BS to STAR-RIS channel $\boldsymbol{H}_{b,r}$, and STAR-RIS to user $u_k$ channel $\boldsymbol{h}_{r,k}$ follow the Rician distribution:
\begin{equation}
\begin{aligned}
    \boldsymbol{H}_{b,r,t} &=\sqrt{\mathcal{L}_{b,r,t}}  
    \left( \sqrt{\frac{Q}{Q+1}}\boldsymbol{H}^{LoS}_{b,r,t}+\sqrt{\frac{1}{Q+1}}\boldsymbol{H}_{b,r,t}^{NLoS} \right),
\end{aligned}
\end{equation}
\begin{equation}
\begin{aligned}
    \boldsymbol{h}_{r,k,t} &=\sqrt{\mathcal{L}_{r,k,t}}  
     \left( \sqrt{\frac{Q}{Q+1}}\boldsymbol{h}^{LoS}_{r,k,t}+\sqrt{\frac{1}{Q+1}}\boldsymbol{h}_{r,k,t}^{NLoS} \right),
\end{aligned}
\end{equation}
where $\mathcal{L}_{b,r,t}$ represents the power domain pathloss, and $Q$ represents the Rician factor. $\boldsymbol{H}_{b,r,t}^{NLoS}$ and $\boldsymbol{h}_{r,k,t}^{NLoS}$ non-line-of-sight (NLoS) component, and follows $\boldsymbol{H}_{b,r,t}^{NLoS}, \boldsymbol{h}_{r,k,t}^{NLoS} \sim \mathcal{C}\mathcal{N}(0,1)$. $\boldsymbol{H}_{b,r,t}^{LoS}$ and $\boldsymbol{h}_{r,u,t}^{LoS}$ represent the LoS channel of BS-to-STAR-RIS and STAR-RIS-to-$u_k$:
\begin{equation}
    \boldsymbol{H}_{b,r,t}^{LoS} = \mathbf{a}_r (\phi^A_B,\psi^A_B) \mathbf{a}_b (\phi^D_B)^H,
\end{equation}
\begin{equation}
    \boldsymbol{h}_{r,k,t}^{LoS} = \mathbf{a}_r (\phi^D_{Rk},\psi^D_{Rk}).
\end{equation}
\par

As presented in 3GPP specification TR 36.873\cite{3GPP}, the path loss $\mathcal{L}$ follows the urban propagation model.
Therefore, in NLoS and LoS channels, the path loss is represented as $\mathcal{L}_{NLoS}$ and $\mathcal{L}_{LoS}$:
\begin{equation}
    \mathcal{L}_{NLoS} = max[\mathcal{L}_{LoS}(d,f_c), \mathcal{L}_{NLoS}(d,f_c)],
\end{equation}
\begin{equation}
\begin{aligned}
    \mathcal{L}_{NLoS}(d,f_c) &= 36.7\log_{10}{d}+22.7 \\
    & +26\log_{10}{f_c}-0.3(z_{R}-1.5),
\end{aligned}
\end{equation}
\begin{equation}
    \mathcal{L}_{LoS}(d,f_c) = 22.0\log_{10}{d}+28.0+20\log_{10}{f_c},
\end{equation}
where $d$ represents the distance between the transmitter and the receiver, and $f_c$ represents the carrier frequency. 
\par

LoS channels cannot be guaranteed between BS and users due to obstacles such as buildings or trees. Therefore, channel $\boldsymbol{h}_{b,k}$ is NLoS channel and follow Rayleigh fading, represented as $\boldsymbol{h}_{b,k} = \sqrt{\mathcal{L}_{b,k}}\boldsymbol{h}_{b,k}^{NLoS}, \boldsymbol{h}_{b,k}^{NLoS} \sim \mathcal{C}\mathcal{N}(0,1)$. 

\subsubsection{\textbf{Signal Model}}
We denote the signals and the active beamforming vectors from the BS towards user $u_k$ as $s_{k,t}$ and $\textbf{w}_{k,t}\in \mathbb{C}^{M \times 1}$, respectively. The signal transmitted at time slot $t$ can be expressed as: $\mathbf{x}_{b,t} = \sum_{k=1}^{K}{\mathbf{w}_{k,t}s_{k,t}}.$
\par

The received signal of $u_k$ with Gaussian noise $n_0$ can be expressed as: 
\begin{equation}
    y_{k,t} = \sum_{i=1}^{K}{(\boldsymbol{h}_{b,k,t} + \boldsymbol{h}_{r,k,t}\Theta_{k,t}\boldsymbol{H}_{b,r,t}) \mathbf{w}_{i,t}s_{i,t}} + n_0,
\end{equation}
\par

With the given received signal of each user, the signal-to-interference-plus-noise ratio (SINR) of user $u_k$ in the reflection and transmission areas can be expressed as:
\begin{equation}
    \gamma_{k,t} = 
    \frac
    {|(\boldsymbol{h}_{b,k,t} + \boldsymbol{h}_{r,k,t}\Theta_{k,t}\boldsymbol{H}_{b,r,t})\mathbf{w}_{k,t}|^2}
    {|(\boldsymbol{h}_{b,k,t} + \boldsymbol{h}_{r,k,t}\Theta_{k,t}\boldsymbol{H}_{b,r,t})\sum_{i=1, i\neq k} ^{K} \mathbf{w}_{i,t}|^2+\sigma^2},
\end{equation}
where $\sigma^2$ denotes the noise power. Therefore, with bandwidth $B$, according to the Shannon's Theorem, the achievable data rate of user $u_k$ is:
\begin{equation} \label{rate}
    R_{k,t} = B\log_{2}{(1+\gamma_{k,t})}.
\end{equation}

\subsection{Problem Formulation}

We aim to maximize the sum-rate for multiple users as defined in \eqref{rate}, by designing the STAR-RIS movement $(\Delta_x, \Delta_y)$ on x-y plane and orientation $\alpha$, and hybrid beamforming of BS and STAR-RIS. Therefore, the optimization problem can be formulated as:
\begin{subequations}
\begin{align}
    \max_{\textbf{w}_k,\Theta_k,\beta,\Delta_{x},\Delta_{y},\alpha} & {\sum_{t=1}^{T} \sum_{k=1}^{K} R_{k,t}} \label{tar} \\
    \mbox{s.t.} \quad \quad
    &\ \theta_{T,n,t},\theta_{R,n,t}\in[0, 2\pi) \label{c1},  \\
    & -x_{max} \leq \Delta_{x,t} \leq x_{max} \label{c6},  \\
    & -y_{max} \leq \Delta_{y,t} \leq y_{max} \label{c7},  \\
    & \sum_{k=1}^{K} ||\textbf{w}_{k,t}||^2 \leq P_{max} \label{c8},  \\
    & (\ref{limit}), (\ref{diagmatrix}), 
\end{align}
\end{subequations}
where constraints \eqref{c6} and \eqref{c7} represent the STAR-RIS movement in each step within a bound. Constraint \eqref{c8} represents that at each time slot $t$, the power of BS is not larger than the max power $P_{max}$.

\section{PPO-based Algorithm for STAR-RIS Joint Deployment and Beamforming Design}

In this section, we present our PPO-based algorithm for dynamic STAR-RIS deployment and hybrid beamforming to maximize the communication sum rate for users. While traditional semidefinite programming methods achieve static optimization, their high computational complexity limits dynamic applications. We adopt deep reinforcement learning (DRL) for environmental adaptability, specifically employing PPO \cite{PPO} due to its stability and efficiency in continuous action space optimization.

The joint STAR-RIS deployment and hybrid beamforming optimization is modeled as a Markov decision process (MDP). At each time step, the DRL agent observes state information from the communication environment. Then the DRL agent determines the optimal STAR-RIS deployment and hybrid beamforming strategy and receives rewards, which in our case is the sum rate of all users.

\begin{algorithm}[t]
\caption{PPO algorithm}\label{PPO process}
\begin{algorithmic}[1]
    \State Initialize $\pi_{0}$ with parameters $\varphi_{0}$ and $V_{0}$ with parameters $\rho_{0}$. Initialize the simulation environment for the wireless communication system.
    \For {$k = 0, 1, 2, \ldots$}
        \While {Batch is not filled}
            \State Observe state $s_t$.
            \State Choose action $a_t$ according current policy $\pi_k$.
            \State Map action $a_t$ to executable actions.
            \State Execute action $a_t$ in environment.
            \State Get reward $r_t$, observe next state $s_{t+1}$.
            \State Store $\{s_t, a_t, r_t, s_{t+1}\}$ in batch.
        \EndWhile
        \State Compute the advantage estimation \eqref{adv2}.
        \State Update $\pi_{k}$ with \eqref{loss function}:
            
        \State Update $V_{k}$ on mean square error:
            $$
            \rho_{k+1} = \arg \min_{\rho} \frac{1}{|\mathcal{D}_k|T}
            \sum_{\tau \in \mathcal{D}_k} \sum_{t=0}^{T}
            \left( 
                V_{\rho}(s_t) - \hat{R}_t
            \right) ^ 2.
            $$
    \EndFor
\end{algorithmic}
\end{algorithm}

\subsection{State, Action, and Reward Design} \label{SAR}
\textbf{State: }We define the observation space $\mathcal{S}$, which includes CSI information of each channel, the user grouping information, and the location of each user. The state $\mathbf{s}_t \in \mathcal{S}$ can be denoted as:
\begin{equation}
    \mathbf{s}_t = \{\boldsymbol{h}_{b,u,t}, \boldsymbol{H}_{b,r,t}, \boldsymbol{h}_{r,u,t}, \mathbf{L}_u, \mathbf{p}_u\},
\end{equation}
where $\boldsymbol{h}_{b,u,t}$ represents the CSI from BS to users, $\boldsymbol{H}_{b,r,t}$ represents the CSI from BS to STAR-RIS, and $\boldsymbol{H}_{r,u,t}$ represents the CSI from the STAR-RIS to users. $\mathbf{L}_u$ represents the users' grouping labels. We denote the grouping label of users at the reflective and transmitting sides as -1 and 1, respectively. And $\mathbf{p}_u$ represents the location of users.

\textbf{Action: }The action space $\mathcal{A}$ includes the movement of STAR-RIS $a_t^{(\Delta_{x}, \Delta_{y})}$, the orientation of STAR-RIS $a_t^{\alpha}$, the passive beamforming on STAR-RIS $(a_t^{\Theta_{R}}, a_t^{\beta_{R,n}}, a_t^{\Theta_{T}})$, and the active beamforming of BS $(a_t^{\Theta_{B}}, a_t^{\beta_{B,m}})$.
The agent will determine the action $\mathbf{a}_t \in \mathcal{A}$ when it gets the state information, and $a_t$ can be denoted as: 
\begin{equation}
    \mathbf{a}_t = \{ a_t^{\Theta_{R}}, a_t^{\Theta_{T}}, a_t^{\beta_{R,n}}, a_t^{\Theta_{B}}, a_t^{\beta_{B,m}}, a_t^{(\Delta_{x}, \Delta_{y})}, a_t^{\alpha}\},
\end{equation}
where each action is in the range $[-1, 1]$, and can be mapped to the actual corresponding executable actions with different mapping functions. $\Theta_{R}$, $\beta_{R, n}$, $\Theta_{B}$, $\beta_{B,m}$, $\Delta_x$, $\Delta_y$, $\alpha$ are continuous in a certain range, therefore they can be generated by multiplying specific coefficients to actions. Given the constraint $\theta_{R} = \theta_{T} \pm \frac{\pi}{2}$ from \eqref{limit}, we define the mapping function of $\theta_{R}$ through its phase relationship with $\theta_{T}$:
\begin{equation}
\theta_{R}=
    \begin{cases}
    \theta_{T}+\frac{\pi}{2},  & \text{if } a_t^{\Theta_{R}} > 0, \\
    \theta_{T}-\frac{\pi}{2},  & \text{otherwise}.
    \end{cases}
\end{equation}
the amplitude of transmitted and reflected signals on element $n$ follows $\beta_{R,n} = \sqrt{1-\beta_{T,n}^2}$, according to \eqref{limit}.

\textbf{Reward: }The optimization objective is to maximize the communication sum-rate, thus we design the reward $\mathbf{r}_t$ as the sum rate for users in the wireless communication network, denoted as:
\begin{equation}
    \mathbf{r}_t = \sum_{k=1}^{K} R_{k,t},
\end{equation}
where $R_{k,t}$ is the rate of $u_k$ at time slot $t$ calculated by \eqref{rate}.

\subsection{PPO Algorithm}

PPO is an on-policy reinforcement learning algorithm, it consists of a critic network $V$ that predicts the value of a state, and an actor-network $\pi$ with parameter $\varphi$ that determines the action to perform. Our proposed PPO-based algorithm operates by having the state $s_t$ from the environment, and determining the action $a_t$ at this time slot, as introduced in Section \ref{SAR}.

The actor is a stochastic policy, and the output action is a probability distribution. The objective of the PPO algorithm is to maximize the expected reward $r$ over a set of trajectories generated by the policy. Therefore, the algorithm updates $\varphi$ each epoch by:
\begin{equation}
    \varphi_{k+1} = \arg \max_{\varphi} \mathbb{E}_{s,a \in \pi_{\varphi_{k}}} \left[ L\left( s, a, \varphi_k, \varphi \right) \right],
\end{equation}
and the loss function $L$ is the clipped advantage estimation with importance sampling, calculated as:
\begin{equation} \label{loss function}
\begin{aligned}
L\left( s, a, \varphi_k, \varphi \right) &= 
\min \left( \frac{\pi_{\varphi}(a|s)}{\pi_{\varphi_{k}}(a|s)} A^{\pi_{\varphi_{k}}}(s,a), \right. \\
&\quad \left. \text{clip} \left( \frac{\pi_{\varphi}(a|s)}{\pi_{\varphi_{k}}(a|s)}, 1-\epsilon, 1+\epsilon \right) A^{\pi_{\varphi_{k}}}(s,a) \right),
\end{aligned}
\end{equation}
in which $\epsilon$ is a hyperparameter that roughly says how far away the new policy is allowed to go from the old, and $A(s, a)$ is the advantage of action $a$ at state $s$.

The estimator of advantage function at time slot $t$ is represented by $\hat{A}_t$, which estimates the advantage of specific action $a$ at state $s$ over other actions in action space $A$. The Generalized Advantage Estimator (GAE)\cite{advantage_function} used in PPO algorithm calculates $\hat{A}_t$ within $T$ time slots, and $T$ can be much smaller than the episode length:
\begin{equation} \label{adv2}
    \hat{A}_{t} = \delta_t + (\gamma \lambda) \delta_{t+1} + \ldots + (\gamma \lambda)^{T-t+1} \delta_{T-1},
\end{equation}
where
\begin{equation}
    \delta_t = r_t + \gamma V(s_{t+1}) - V(s_t),
\end{equation}
where $t$ specifies the time index in $[0, T]$, within a given length-$T$ trajectory $\tau$ segment, $\lambda$ is the GAE parameter and $\gamma$ is the discount factor. $V$ represents the critic network predict function, which outputs the state value. The overall training process is shown in \textbf{Algorithm \ref{PPO process}}.

\section{Simulation Results}
In this section, we provide numerical simulations to verify the efficiency of our proposed dynamic deployment scheme and approach. 

\subsection{Simulation Settings}
Three deployment schemes are compared, including
\textbf{1) Proposed:} The location and orientation of the STAR-RIS, as well as hybrid beamforming, are jointly optimized by the proposed algorithm based on PPO;
\textbf{2) Fixed STAR-RIS:} The location and orientation are fixed, and only the hybrid beamforming is trained using the proposed PPO-based algorithm;
\textbf{3) No STAR-RIS:} The wireless communication network is not assisted by STAR-RIS.

The default parameters for the simulation implementation are listed in Table \ref{table3}. 

\begin{table}[t]
    \centering
    \caption{Parameter settings} \label{table3}
    \begin{tabular}{l|c}
    \hline
         Parameter                                          &   value \\  
         \hline
         BS antenna number $M$                    &   4 \\
         STAR-RIS element number $N$                        &   15 \\
         STAR-RIS element number on each row $N_x$          &   5 \\
         number of users $K$                                &   6 \\
         carrier frequency $f_c$                            &   5GHz \\
         bandwidth $B$                                      &   1MHz \\
         Rician factor $Q$                                  &   5dB \\
         BS position $(x_{B},y_{B},x_{B})$                  &   $(0, 1000, 15)$ \\
         STAR-RIS maximum movement $x_{max}, y_{max}$       &   15m \\
         user maximum movement                              &   5m \\
         STAR-RIS initial position $(x_{R},y_{R},x_{R})$    &   $(0, 0, 20)$ \\
         \hline
         discount factor $\gamma$               &   0.99 \\
         GAE parameter $\lambda$                &   0.95 \\
         clipping threshold $\epsilon$          &   0.2 \\
         batch size                             &   8192 \\
         steps per episode                      &   30 \\
         number of layers for actor and critic  &   3 \\
         hidden layer neurons                   &   64 \\
         learning rate $lr$                     &   $3 \times 10^{-3}$ \\
         optimizer                              &   `AdamW' \\
    \hline
    \end{tabular}
\end{table}

\subsection{Results and Analysis}
\begin{figure}
\centering
\includegraphics[width=0.5\textwidth]{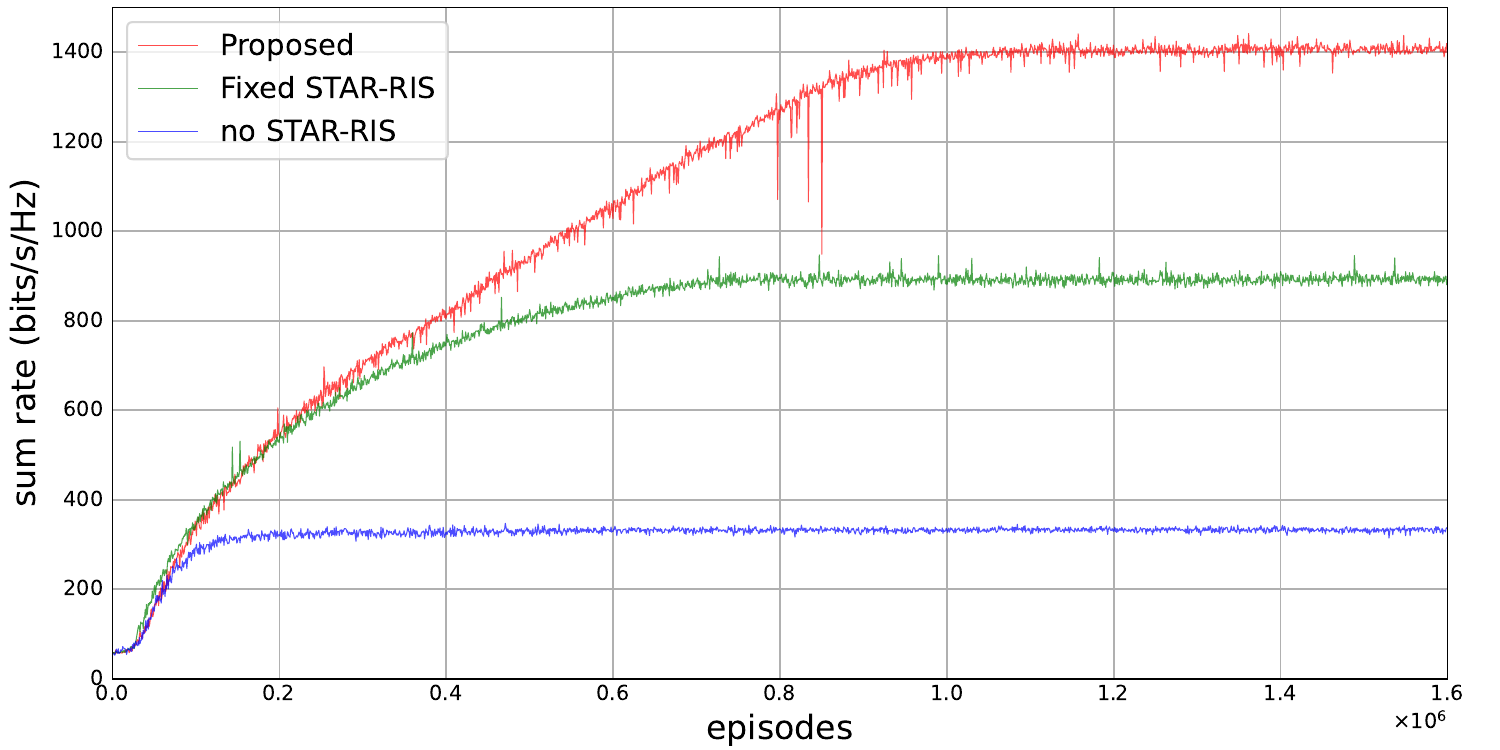}
\caption{Sum rate comparison between deployed STAR-RIS, fixed STAR-RIS, and no STAR-RIS assisted wireless network.}
\label{result1}
\end{figure}

Figure~\ref{result1} illustrates the convergence curves of the three deployment schemes in terms of the sum rate. 
The proposed PPO-based algorithm converges reliably across all settings, demonstrating strong adaptability and learning efficiency. 
In particular, the dynamic deployment scheme consistently outperforms both the fixed STAR-RIS and RIS-free baselines. 
This gain can be attributed to the joint optimization of location and orientation, which dynamically adjusts user grouping and avoids mode congestion (e.g., excessive users in either transmission or reflection region). 
This adaptability is especially beneficial under user mobility and varying channel conditions. 

\begin{figure}[]
\centering
\includegraphics[width=0.5\textwidth]{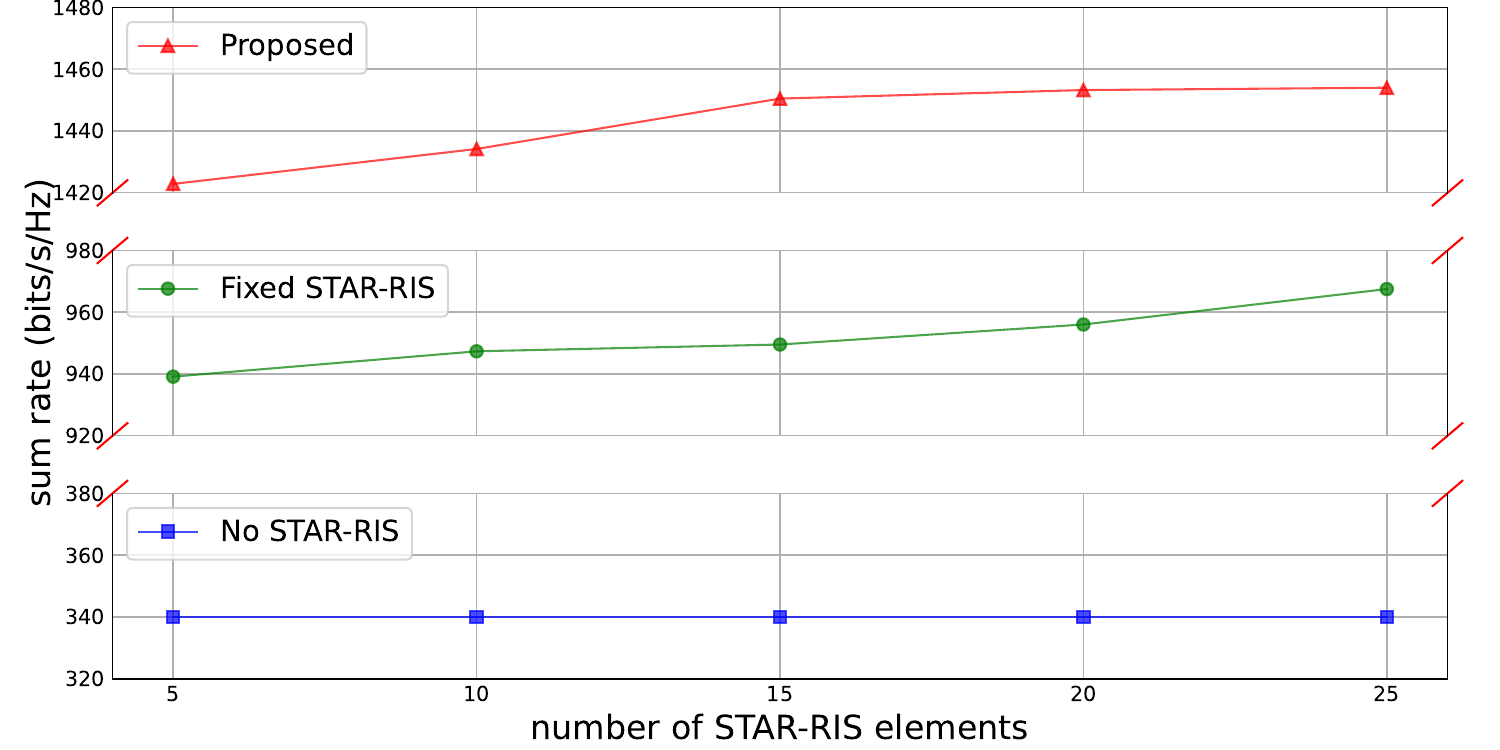}
\caption{Performances versus the number of STAR-RIS elements}
\label{result2}
\end{figure}

We further evaluate the performance of three scenarios across varying numbers of STAR-RIS elements. As shown in Figure \ref{result2}, the sum rate of all users consistently increased with the number of STAR-RIS elements, aligning with the theoretical expectations of diversity gain. As the number of elements increases, the proposed scheme undergoes a flattening state in sum rate. This phenomenon may be attributed to the limited spatial diversity and channel correlation in smaller networks, where the capacity improvement from additional elements saturates. 

\begin{figure}[]
\centering
\includegraphics[width=0.45\textwidth]{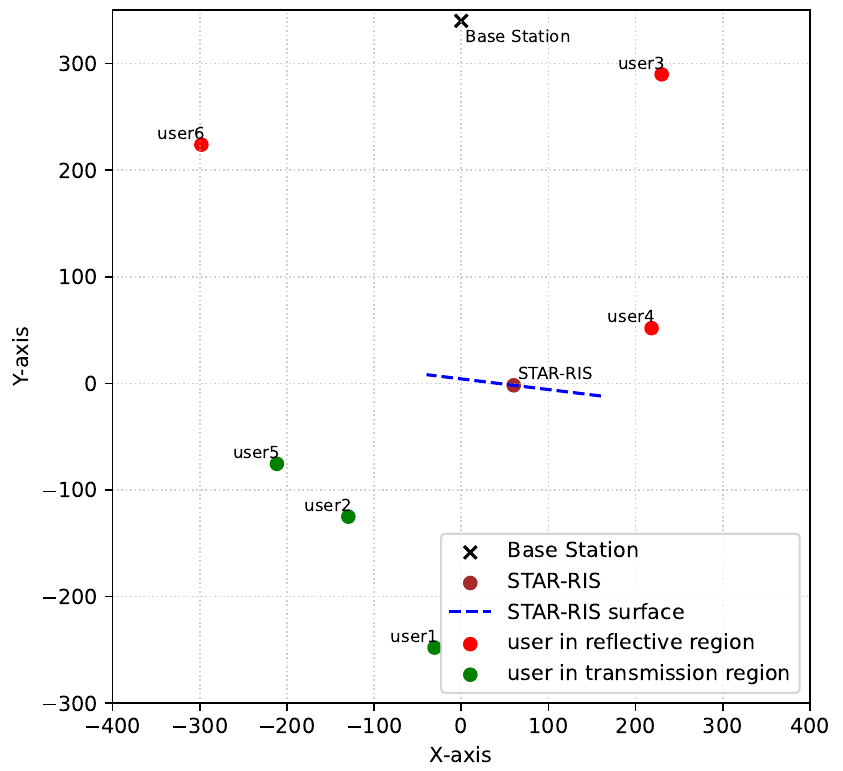}
\caption{STAR-RIS deployment strategy example}
\label{scenario}
\end{figure}

Figure \ref{scenario} illustrates the optimal position and orientation of the STAR-RIS determined by the proposed algorithm in a specific time interval throughout the service episode. The figure shows that the algorithm strategically positions the STAR-RIS near the geometric center of the user distribution, to minimize cascaded channel losses for all users. Furthermore, the deployment effectively divides users into balanced reflective and transmissive groups. This partitioning prevents overcrowding in a single beamforming mode, which would otherwise limit design flexibility and reduce total communication rates. These findings validate the ability of the proposed algorithm to dynamically adapt STAR-RIS configurations for performance optimization in evolving wireless environments.




\section{Conclusion}
This paper proposes a dynamic control framework for aerial STAR-RIS-aided MISO systems. Unlike prior works that overlook/oversimplify deployment-induced user grouping, we explicitly model how STAR-RIS placement affects transmission/reflection regions. We formulate a joint optimization and develop a PPO-based algorithm to adaptively balance grouping and beamforming in dynamic environments. Simulation results have demonstrated that the proposed dynamic deployment scheme achieves a 57.1\% and 285\% improvement in sum rate compared to fixed deployment and RIS-free scenarios, respectively, which clearly highlighted the superior performance of the proposed scheme in addressing the challenges of realistic dynamic mobile communication systems.


The proposed scheme can be seamlessly extended to multi-BS scenarios through a multi-agent framework where each BS-located agent independently implements the algorithm while maintaining inter-agent coordination. Additionally, methods like clustered beamforming can be integrated to handle large-scale user scenarios. Our MDP formulation and RL solution address the core optimization challenge, and the exploration of other RL methods remains future work.


\bibliographystyle{IEEEtran}
\bibliography{Ref_STAR}

@string{icc= "Proc. IEEE Int. Conf. on Communications (ICC)"}

@ARTICLE{aung2023aerial,
  author={Aung, Pyae Sone and Nguyen, Loc X. and Tun, Yan Kyaw and Han, Zhu and Hong, Choong Seon},
  journal={IEEE Wireless Communications Letters}, 
  title={Aerial {STAR-RIS} Empowered {MEC}: A {DRL} Approach for Energy Minimization}, 
  year={2024},
  volume={13},
  number={5},
  pages={1409-1413},
  doi={10.1109/LWC.2024.3372623}}

@ARTICLE{wang2023average,
  author={Wang, Peiyu and Wang, Hong and Fu, Yibo},
  journal={IEEE Communications Letters}, 
  title={Average Rate Maximization for Mobile {STAR-RIS}-Aided {NOMA} System}, 
  year={2023},
  volume={27},
  number={5},
  pages={1362-1366},
  doi={10.1109/LCOMM.2023.3261442}}

@article{xiao2024star,
  title={{STAR-RIS} Enhanced {UAV}-Enabled {MEC} Networks with Bi-Directional Task Offloading},
  author={Xiao, Han and Hu, Xiaoyan and Mu, Pengcheng and Zhang, Weile and Wang, Wenjie and Wong, Kai-Kit and Yang, Kun},
  journal={arXiv preprint arXiv:2401.05725},
  year={2024}
}

@ARTICLE{su2023joint,
  author={Su, Yuhua and Pang, Xiaowei and Lu, Weidang and Zhao, Nan and Wang, Xianbin and Nallanathan, Arumugam},
  journal={IEEE Transactions on Vehicular Technology}, 
  title={Joint Location and Beamforming Optimization for {STAR-RIS} Aided {NOMA-UAV} Networks}, 
  year={2023},
  volume={72},
  number={8},
  pages={11023-11028},
  doi={10.1109/TVT.2023.3261324}}

@ARTICLE{zhang2022joint,
  author={Zhang, Qin and Zhao, Yang and Li, Hai and Hou, Shujuan and Song, Zhengyu},
  journal={IEEE Wireless Communications Letters}, 
  title={Joint Optimization of {STAR-RIS} Assisted {UAV} Communication Systems}, 
  year={2022},
  volume={11},
  number={11},
  pages={2390-2394},
  doi={10.1109/LWC.2022.3204353}}

@ARTICLE{gao2022joint,
  author={Gao, Qiling and Liu, Yuanwei and Mu, Xidong and Jia, Min and Li, Dongbo and Hanzo, Lajos},
  journal={IEEE Transactions on Communications}, 
  title={Joint Location and Beamforming Design for {STAR-RIS} Assisted {NOMA} Systems}, 
  year={2023},
  volume={71},
  number={4},
  pages={2532-2546},
  doi={10.1109/TCOMM.2023.3247753}}

@ARTICLE{2inSTARZhong,
  author={Mu, Xidong and Liu, Yuanwei and Guo, Li and Lin, Jiaru and Al-Dhahir, Naofal},
  journal={IEEE Transactions on Wireless Communications}, 
  title={Exploiting Intelligent Reflecting Surfaces in NOMA Networks: Joint Beamforming Optimization}, 
  year={2020},
  volume={19},
  number={10},
  pages={6884-6898}}

@ARTICLE{3inSTARZhong,
  author={ElMossallamy, Mohamed A. and Zhang, Hongliang and Song, Lingyang and Seddik, Karim G. and Han, Zhu and Li, Geoffrey Ye},
  journal={IEEE Transactions on Cognitive Communications and Networking}, 
  title={Reconfigurable Intelligent Surfaces for Wireless Communications: Principles, Challenges, and Opportunities}, 
  year={2020},
  volume={6},
  number={3},
  pages={990-1002}}

@ARTICLE{12inSTARZhong,
  author={Mu, Xidong and Liu, Yuanwei and Guo, Li and Lin, Jiaru and Schober, Robert},
  journal={IEEE Transactions on Wireless Communications}, 
  title={Simultaneously Transmitting and Reflecting {(STAR) RIS} Aided Wireless Communications}, 
  year={2022},
  volume={21},
  number={5},
  pages={3083-3098}}

@article{14inSTARZhong,
  title={Dynamic control of electromagnetic wave propagation with the equivalent principle inspired tunable metasurface},
  author={Zhu, Bo O and Chen, Ke and Jia, Nan and Sun, Liang and Zhao, Junming and Jiang, Tian and Feng, Yijun},
  journal={Scientific reports},
  volume={4},
  number={1},
  pages={4971},
  year={2014},
  publisher={Nature Publishing Group UK London}
}

@article{usermovements,
title={Machine Learning Empowered Trajectory and Passive Beamforming Design in {UAV-RIS} Wireless Networks},
volume={39}, 
ISSN={1558-0008}, 
abstractNote={A novel framework is proposed for integrating reconfigurable intelligent surfaces (RIS) in unmanned aerial vehicle (UAV) enabled wireless networks, where an RIS is deployed for enhancing the service quality of the UAV. Non-orthogonal multiple access (NOMA) technique is invoked to further improve the spectrum efficiency of the network, while mobile users (MUs) are considered as roaming continuously. The energy consumption minimizing problem is formulated by jointly designing the movement of the UAV, phase shifts of the RIS, power allocation policy from the UAV to MUs, as well as determining the dynamic decoding order. A decaying deep Q-network (D-DQN) based algorithm is proposed for tackling this pertinent problem. In the proposed D-DQN based algorithm, the central controller is selected as an agent for periodically observing the state of UAV-enabled wireless network and for carrying out actions to adapt to the dynamic environment. In contrast to the conventional DQN algorithm, the decaying learning rate is leveraged in the proposed D-DQN based algorithm for attaining a tradeoff between accelerating training speed and converging to the local optimal. Numerical results demonstrate that: 1) In contrast to the conventional Q-learning algorithm, which cannot converge when being adopted for solving the formulated problem, the proposed D-DQN based algorithm is capable of converging with minor constraints; 2) The energy dissipation of the UAV can be significantly reduced by integrating RISs in UAV-enabled wireless networks; 3) By designing the dynamic decoding order and power allocation policy, the RIS-NOMA case consumes 11.7 less energy than the RIS-OMA case.}, 
number={7}, 
journal={IEEE Journal on Selected Areas in Communications}, 
author={Liu, Xiao and Liu, Yuanwei and Chen, Yue}, 
year={2021}, 
pages={2042–2055} }

@article{multihop, 
   title={Multi-Hop {RIS}-Empowered Terahertz Communications: A {DRL}-Based Hybrid Beamforming Design}, 
   volume={39}, 
   ISSN={1558-0008}, 
   abstractNote={Wireless communication in the TeraHertz band (0.1-10 THz) is envisioned as one of the key enabling technologies for the future sixth generation (6G) wireless communication systems scaled up beyond massive multiple input multiple output (Massive-MIMO) technology. However, very high propagation attenuations and molecular absorptions of THz frequencies often limit the signal transmission distance and coverage range. Benefited from the recent breakthrough on the reconfigurable intelligent surfaces (RIS) for realizing smart radio propagation environment, we propose a novel hybrid beamforming scheme for the multi-hop RIS-assisted communication networks to improve the coverage range at THz-band frequencies. Particularly, multiple passive and controllable RISs are deployed to assist the transmissions between the base station (BS) and multiple single-antenna users. We investigate the joint design of digital beamforming matrix at the BS and analog beamforming matrices at the RISs, by leveraging the recent advances in deep reinforcement learning (DRL) to combat the propagation loss. To improve the convergence of the proposed DRL-based algorithm, two algorithms are then designed to initialize the digital beamforming and the analog beamforming matrices utilizing the alternating optimization technique. Simulation results show that our proposed scheme is able to improve 50 more coverage range of THz communications compared with the benchmarks. Furthermore, it is also shown that our proposed DRL-based method is a state-of-the-art method to solve the NP-hard beamforming problem, especially when the signals at RIS-assisted THz communication networks experience multiple hops.}, 
   number={6}, 
   journal={IEEE Journal on Selected Areas in Communications}, 
   author={Huang, Chongwen and Yang, Zhaohui and Alexandropoulos, George C. and Xiong, Kai and Wei, Li and Yuen, Chau and Zhang, Zhaoyang and Debbah, Mérouane}, 
   year={2021}, 
   pages={1663–1677} }

@article{ZhongRuikang, 
    title={Hybrid Reinforcement Learning for {STAR-RIS}s: A Coupled Phase-Shift Model Based Beamformer}, 
    volume={40}, 
    ISSN={1558-0008}, 
    abstractNote={A simultaneous transmitting and reflecting reconfigurable intelligent surface (STAR-RIS) assisted multi-user downlink multiple-input single-output (MISO) communication system is investigated. In contrast to the existing ideal STAR-RIS model assuming an independent transmission and reflection phase-shift control, a practical coupled phase-shift model is considered. Then, a joint active and passive beamforming optimization problem is formulated for minimizing the long-term transmission power consumption, subject to the coupled phase-shift constraint and the minimum data rate constraint. Despite the coupled nature of the phase-shift model, the formulated problem is solved by invoking a hybrid continuous and discrete phase-shift control policy. Inspired by this observation, a pair of hybrid reinforcement learning (RL) algorithms, namely the hybrid deep deterministic policy gradient (hybrid DDPG) algorithm and the joint DDPG and deep-Q network (DDPG-DQN) based algorithm are proposed. The hybrid DDPG algorithm controls the associated high-dimensional continuous and discrete actions by relying on the hybrid action mapping. By contrast, the joint DDPG-DQN algorithm constructs two Markov decision processes (MDPs) relying on an inner and an outer environment, thereby amalgamating the two agents to accomplish a joint hybrid control. Simulation results demonstrate that the STAR-RIS has superiority over other conventional RISs in terms of its energy consumption. Furthermore, both the proposed algorithms outperform the baseline DDPG algorithm, and the joint DDPG-DQN algorithm achieves a superior performance, albeit at an increased computational complexity.}, 
    number={9}, 
    journal={IEEE Journal on Selected Areas in Communications}, 
    author={Zhong, Ruikang and Liu, Yuanwei and Mu, Xidong and Chen, Yue and Wang, Xianbin and Hanzo, Lajos}, 
    year={2022}, 
    pages={2556–2569} }

@article{PPO, 
 title={Proximal Policy Optimization Algorithms}, 
 abstractNote={We propose a new family of policy gradient methods for reinforcement learning, which alternate between sampling data through interaction with the environment, and optimizing a “surrogate” objective function using stochastic gradient ascent. Whereas standard policy gradient methods perform one gradient update per data sample, we propose a novel objective function that enables multiple epochs of minibatch updates. The new methods, which we call proximal policy optimization (PPO), have some of the benefits of trust region policy optimization (TRPO), but they are much simpler to implement, more general, and have better sample complexity (empirically). Our experiments test PPO on a collection of benchmark tasks, including simulated robotic locomotion and Atari game playing, and we show that PPO outperforms other online policy gradient methods, and overall strikes a favorable balance between sample complexity, simplicity, and wall-time.}, 
 note={arXiv:1707.06347 [cs]}, 
 number={arXiv:1707.06347}, 
 publisher={arXiv}, 
 author={Schulman, John and Wolski, Filip and Dhariwal, Prafulla and Radford, Alec and Klimov, Oleg}, 
 year={2017}, 
 month={Aug} }

@article{advantage_function, 
 title={High-Dimensional Continuous Control Using Generalized Advantage Estimation}, 
 abstractNote={Policy gradient methods are an appealing approach in reinforcement learning because they directly optimize the cumulative reward and can straightforwardly be used with nonlinear function approximators such as neural networks. The two main challenges are the large number of samples typically required, and the difficulty of obtaining stable and steady improvement despite the nonstationarity of the incoming data. We address the first challenge by using value functions to substantially reduce the variance of policy gradient estimates at the cost of some bias, with an exponentially-weighted estimator of the advantage function that is analogous to TD(lambda). We address the second challenge by using trust region optimization procedure for both the policy and the value function, which are represented by neural networks. Our approach yields strong empirical results on highly challenging 3D locomotion tasks, learning running gaits for bipedal and quadrupedal simulated robots, and learning a policy for getting the biped to stand up from starting out lying on the ground. In contrast to a body of prior work that uses hand-crafted policy representations, our neural network policies map directly from raw kinematics to joint torques. Our algorithm is fully model-free, and the amount of simulated experience required for the learning tasks on 3D bipeds corresponds to 1-2 weeks of real time.}, 
 note={arXiv:1506.02438 [cs]}, 
 number={arXiv:1506.02438}, 
 publisher={arXiv}, 
 author={Schulman, John and Moritz, Philipp and Levine, Sergey and Jordan, Michael and Abbeel, Pieter}, 
 year={2018}, 
 month={Oct} }

@techreport{3GPP,
    title = {Study on {3D} Channel Model for {LTE}},
    institution = {3GPP},
    year = {2018},
    number = {document 3GPP TR 36.873 Release 12}
}

@article{orientation, title={Power Control and Passive Beamforming for the {STAR-RIS} With Rotatable Angles}, volume={73}, rights={https://ieeexplore.ieee.org/Xplorehelp/downloads/license-information/IEEE.html}, ISSN={0018-9545, 1939-9359}, DOI={10.1109/TVT.2024.3369617}, abstractNote={Simultaneously transmitting and reﬂecting RISs (STAR-RISs) can serve users positioned on either side of the surface by means of transmitting and reﬂecting the incident signal. The deployment angle of the STAR-RIS is pivotal to the gain of the received and transmitted signals, and the STARRIS with rotatable angles has the potential to boost the spectral efﬁciency (SE). In this paper, the power allocation, transmission and reﬂection coefﬁcients (TARCs), and deployment angle of the STAR-RIS is jointly optimized to maximize the SE. However, since the non-convex of the objective function and the manual coupling of variables, it is challenging to solve this problem. To this end, we propose a low-complexity algorithm, where the deployment angle of the STAR-RIS is obtained by using deep learning (DL) according to channels from the BS to STAR-RIS and from the STAR-RIS to users. Then, the power allocation and TARCs are iteratively optimized in an alternating manner. Numerical results reveal that the proposed algorithm can effectively achieve superior system’s SE.}, number={8}, journal={IEEE Transactions on Vehicular Technology}, author={Wang, Jun-Bo and Zhu, Bingqian and Pan, Yijin and Chen, Yijian and Yu, Hongkang and Tang, Anzheng and Wang, Jiangzhou}, year={2024}, pages={12121–12125}, language={en} }

@ARTICLE{UAVenergy,
  author={Aung, Pyae Sone and Nguyen, Loc X. and Tun, Yan Kyaw and Han, Zhu and Hong, Choong Seon},
  journal={IEEE Wireless Communications Letters}, 
  title={Aerial {STAR-RIS} Empowered {MEC}: A {DRL} Approach for Energy Minimization}, 
  year={2024},
  volume={13},
  number={5},
  pages={1409-1413},
  doi={10.1109/LWC.2024.3372623}}

@article{orideploy,
  title={Heuristic Solution to Joint Deployment and Beamforming Design for {STAR-RIS} Aided Networks},
  author={Yan, Bai and Zhao, Qi and Zhang, Jin and Zhang, J Andrew},
  journal={arXiv preprint arXiv:2404.09149},
  year={2024}
}

@ARTICLE{ChannelEstimation,
  author={Wu, Chenyu and You, Changsheng and Liu, Yuanwei and Gu, Xuemai and Cai, Yunlong},
  journal={IEEE Communications Letters}, 
  title={Channel Estimation for {STAR-RIS}-Aided Wireless Communication}, 
  year={2022},
  volume={26},
  number={3},
  pages={652-656},
  doi={10.1109/LCOMM.2021.3139198}}

@ARTICLE{tutorial,
  author={Chen, Hui and Sarieddeen, Hadi and Ballal, Tarig and Wymeersch, Henk and Alouini, Mohamed-Slim and Al-Naffouri, Tareq Y.},
  journal={IEEE Communications Surveys and Tutorials}, 
  title={A Tutorial on Terahertz-Band Localization for 6G Communication Systems}, 
  year={2022},
  volume={24},
  number={3},
  pages={1780-1815},
  doi={10.1109/COMST.2022.3178209}}

@INPROCEEDINGS{nonperfectCSI,
  author={Saglam, Baturay and Gurgunoglu, Doga and Kozat, Suleyman S.},
  booktitle={2023 IEEE International Conference on Communications Workshops (ICC Workshops)}, 
  title={Deep Reinforcement Learning Based Joint Downlink Beamforming and {RIS} Configuration in {RIS}-Aided {MU-MISO} Systems Under Hardware Impairments and Imperfect {CSI}}, 
  year={2023},
  volume={},
  number={},
  pages={66-72},
  doi={10.1109/ICCWorkshops57953.2023.10283517}}

@article{YAN2022109725,
title = {Fitness landscape analysis and niching genetic approach for hybrid beamforming in {RIS}-aided communications},
journal = {Applied Soft Computing},
volume = {131},
pages = {109725},
year = {2022},
issn = {1568-4946},
doi = {https://doi.org/10.1016/j.asoc.2022.109725},
author = {Bai Yan and Qi Zhao and Mengke Li and Jin Zhang and J. Andrew Zhang and Xin Yao}
}

@article{new_dep, title={Joint Location and Beamforming Design for Energy Efficient STAR-RIS-Aided ISAC Systems}, volume={29}, ISSN={1558-2558}, DOI={10.1109/LCOMM.2024.3503754}, abstractNote={In this letter, we investigate a simultaneously transmitting and reflecting reconfigurable intelligent surface (STAR-RIS) aided integrated sensing and communication (ISAC) system. Our object is to maximize the energy efficiency (EE) of the ISAC system by jointly optimizing the beamforming at the base station (BS), the transmitting/reflecting coefficient matrices of the STAR-RIS, as well as the deployment location of the STAR-RIS. Since the formulated problem is non-convex, we propose an alternative optimization scheme by decomposing the original problem into three sub-problems, where the sub-problems are solved based on semidefinite relaxation (SDR) and fractional programming. Simulation results demonstrate that the STAR-RIS aided ISAC system achieves a higher EE than conventional RIS scheme, and the location optimization for STAR-RIS can significantly improve the system EE. Furthermore, the EE declines rapidly when the radar beampattern gain threshold increases to a larger value.}, number={1}, journal={IEEE Communications Letters}, author={Zhang, Qin and Wu, Han and Li, Hai and Song, Zhengyu and Hou, Shujuan}, year={2025}, month=jan, pages={140–144} }

@article{2025static, title={UAV Onboard STAR-RIS Service Enhancement Mechanism Based on Deep Reinforcement Learning}, volume={25}, rights={https://creativecommons.org/licenses/by/4.0/}, ISSN={1424-8220}, DOI={10.3390/s25061943}, abstractNote={UAVs and reconfigurable intelligent surfaces (RISs) have emerged as promising solutions to enhance communication coverage and performance. However, existing studies primarily focus on optimizing the amplitude and phase shift of a STAR-RIS without considering the impact of varying UAV hovering angles on signal reflection and transmission. In this paper, we propose a novel STAR-RIS-assisted UAV service enhancement mechanism that dynamically adjusts reflection/transmission regions based on the real-time user distribution, significantly improving the channel quality for both edge and occluded users. This work is the first to jointly optimize the phase and amplitude of the STAR-RIS, the UAV flight trajectory, and the hovering angle, addressing the critical challenge of co-channel interference caused by dynamically partitioned service areas. The complex optimization problem is decomposed into subproblems, where the UAV flight trajectory is optimized using the Chained Lin–Kernighan (CLK) algorithm and the STAR-RIS parameters and UAV hovering angle are optimized using the TD3 algorithm. The experimental results show that the proposed mechanism effectively reduces the system service time and user transmission time, outperforming traditional methods.}, number={6}, journal={Sensors}, author={Yan, Junjie and Xu, Yichen and Yuan, Haohao and Xue, Chunhua}, year={2025}, month=mar, pages={1943}, language={en} }

\end{document}